\documentclass[pre,aps,showpacs,twocolumn,superscriptaddress,10pt,a4paper]{revtex4}
\usepackage{graphicx}

\begin{document}

\title{\bf Incipience of quantum chaos in the Jahn-Teller model}

\author{Eva Majern\'{\i}kov\'a}
\email{fyziemar@savba.sk} \affiliation{Institute of Physics,
Slovak Academy of Sciences, D\'ubravsk\'a cesta 9, SK-84 511
Bratislava, Slovak Republic} \affiliation{Department of
Theoretical Physics, Palack\'y University, T\v r. 17. Listopadu
50, CZ-77207 Olomouc, Czech Republic}

\author{Serge Shpyrko}
\email[e-mail:]{serge_shp(at)yahoo.com}
\altaffiliation{On leave from Institute for Nuclear Research,
Ukrainian Academy of Sciences, pr.Nauki 47 Kiev, Ukraine}
\affiliation{Department of Theoretical Physics, Palack\'y
University, T\v r. 17. Listopadu 50, CZ-77207 Olomouc, Czech
Republic}

\received{26 September 2005} \revised{16 February 2006}
\published{24 June 2006}

\begin{abstract}

We studied complex spectra of a two-level electron system coupled
to two phonon (vibron) modes represented by the E$\otimes$e
Jahn-Teller model. For particular rotation quantum numbers we
found a coexistence of up to three regions of the spectra, (i) the
dimerized region of long-range ordered (extended) pairs of
oscillating levels, (ii) the short-range-ordered (localized) "kink
lattice" of avoiding levels, and (iii) the intermediate region of
kink nucleation with variable range of ordering. This structure
appears above a certain critical line as a function of interaction
strength. The level clustering and level avoiding generic patterns
reflect themselves in several intermittent regions between up to
three branches of spectral entropies. Linear scaling behavior of
the widths of level curvature probability distributions provides
the conventionally adopted indication for the presence of quantum
chaos. Level spacing probability distributions show peculiarities
of the partial (for fixed quantum angular momentum) as well as of
the cumulative (all angular momenta) case. The clustering of
levels with two and three dominant spacings at fixed angular
momenta causes notable deviations of the cumulative distributions
from the Poissonian one.
\end{abstract}


\maketitle

\section{Introduction}

Generally, a multitude of avoided crossings in complex spectra of
a quantum system is considered as the signature of chaos which
appears in a quasiclassical limit of the respective nonlinear
many-body system with repulsive interactions
\cite{Eckhardt:1988,Nakamura:1993}.
 In solid-state physics, quantum chemistry, or  quantum optics
investigations of complex excited energy spectra of models with
two electron levels coupled with one or more boson modes are of
special relevance for understanding experimental -- e.g., optical
or transport-properties.

 Numerical investigations of the excited spectra of certain two-level
one- and two-boson systems [e.g., exciton
\cite{Wagner:1992,Herfort:2001} and E$\otimes $e Jahn-Teller (JT)
model \cite{Kong:1990}] show the existence of level avoidings and
related highly excited "exotic" (localized) states. Signatures of
chaos in two-level boson systems remain objects of interest since
the early studies by Hamiltonian methods
\cite{Belobrov:1976,Milonni:1983,Furuya:1998,Graham:1984,Eidson:1986,
Esser:1994}. Later on statistical methods based on the random
matrix theory -- e.g., level spacing and level curvature
probability distributions -- were applied mostly for
one-phonon-mode models. The level spacing distributions show
irregular behavior between the limiting Poisson (regular) and
Wigner (fully chaotic) distributions
\cite{Lewen:1991,Cibils:1995,Steib:1998,Graham:1986}.
 Specifically, among the level spacing distributions of the
two-level one-phonon model
\cite{Lewen:1991,Cibils:1995,Steib:1998} there appeared M-shaped
distributions with two symmetric peaks for two dominant spacings.
Such distributions are a signature of level clustering
(dimerization).

 The most familiar representative of the two-level two-phonon models
 is the E$\otimes $e JT model \cite{OBrien:1964} with linear coupling to phonon
(vibron) modes of different parity against reflection. This model
is a prototype for phonons removing the
 degeneracy of electron levels by an interaction with the antisymmetric
phonon mode
 and {\it tunneling between the levels assisted by the symmetric phonon
 mode}  (in the one-phonon case the tunneling is purely resonant).
  The presence of two-phonon modes with equal coupling strengths imposes
  additional rotational symmetry to the model. Differences between the
one-mode two-level phonon model and two-mode JT models become
evident when one compares their dimensionality after elimination
of the electron (level) degrees of freedom: while in the first
case the reduction results in a one-dimensional nonlinear (with
self-interaction) quantum oscillator, in the two-mode JT cases the
elimination yields a two-dimensional quantum oscillator with
nonlinear coupling of its components.  While in the one-boson-mode
case one finds correlations of two sets of excited states, in the
two-boson-mode case mixing of up to three sets takes place
depending on the range of parameters. As we show, this leads to
the appearance of a third peak in the level spacing distribution
(trimerization) for the respective range of parameters.

 Recently, Yamasaki {\it et al.} \cite{Nakamura:2003} for the first 
time investigated
the possibility of chaos in spectra of the E$\otimes$e JT model.
 Their analysis was based on the approximation of the Hamiltonian by
 the
 adiabatic "Mexican hat" potential. Additionally the Hamiltonian
 was supplemented by
 an explicit term  with nonlinear mode coupling of a
trigonal symmetry in order to simulate the effects of fluctuations
and nonintegrability. Thus, nonlinearity was included via mode
coupling in addition to the mean-field bare part of the
Hamiltonian. The authors concluded that quantum chaos reflects
itself in the Wigner-type level spacing distribution as a
consequence of the said nonlinearity of the Hamiltonian.
Meanwhile, for the linear part an absence of such patterns was
stated. Those conclusions were confirmed at the classical level as
well.
 It is to be noted here that the
way of passing to a semiclassical approximation in spin-boson
systems is not unique and presents an essential ambiguity from
different possible ways of decoupling \cite{Graham:1984}.
 Different ways of performing a
semiclassical approximation are known to lead to different answers
concerning the chaotic behavior of the system. This ambiguity
means that a classical analog of such a model is not well defined
and cannot be a reliable object in exploring the quantum chaos
issues.

In this paper we investigate thecharacteristics of excited (quasi
continuum) spectra of E$\otimes$e JT model. The present approach
differs from that by Yamasaki {\it et al.} in the following: (a)
we {\it do not} introduce explicit nonlinearity into the initial
Hamiltonian. However, its SU(2) symmetry involves an {\it
intrinsic nonlinearity} which is revealed by exact elimination of
the electronic degrees of freedom \cite{Majernikova:2003}.
 (b) We start from the E$\otimes (b_1+b_2)$ model
 with different coupling strengths for both modes.
 The rotation-symmetric E$\otimes$e model represents its
particular case with equal coupling constants and thus with the
symmetry of a higher (rotational) symmetry group (the difference
of the coupling constants in realistic systems is likely to be
caused, for example, by spatial anisotropy of crystals). The
importance of starting from a more general situation is evident
from the nature of quantum fluctuations. Namely, the variational
approach to the E$\otimes (b_1+b_2)$ model used for the
calculation of the ground state \cite{Majernikova:2003} yielded
the largest deviations just for the rotation-symmetric case. In
the ground state the abrupt change of energy at equal couplings is
an artefact of the adiabatic approximation: the energy region of
the quantum E$\otimes $e model is situated
 within a smeared border around $\beta/\alpha=1$ ($\alpha$ and
 $\beta$ being the coupling strengths of the antisymmetric and symmetric
 modes, respectively)
between the "selftrapping" ($\alpha>\beta$) and "tunneling"
($\alpha <\beta$) part of the ground state of the model with
broken rotational symmetry. In the smeared transition region both
phases coexist and are quantum correlated (entangled) via
nonlinear correlations between the modes. There the
phonon-assisted (by the symmetric mode) tunneling contribution to
the energy (from the admixture of two reflection symmetric levels
in the excitation reflection Ansatz for the variational wave
function\cite{Majernikova:2003}) causes an essential decrease of
the ground state energy because of changing parity of the wave
function by the reflection operator and results in the vanishing
of the Ham factor. The phase transition at $\alpha=\beta $ to the
rotational symmetry phase provides a representation by the
additional rotational quantum number $j$. We will show that at a
given $j$ in E$\otimes $e model analogous phonon assisted
tunneling between adjacent excited levels occurs and is
responsible for the flip (kink) between the levels [this mechanism
can be well recognized, e.g., in Fig.1(b)]. This nonlinear effect
causes a mixture of the adjacent (even and odd) levels at stronger
couplings and leads to nucleation of the new
 kink lattice phase when increasing the coupling constant.
Naturally, the tunneling mechanism due to short-range quantum
fluctuations at large $j$ is then the origin of the peculiar
irregular behavior of related level spacing distributions and does
not allow for full developing of Wigner chaos.
  The phonon-assisted tunneling and nucleation of the kink lattice phase
can also be demonstrated  within the formalism of a Calogero-Moser
gas of pseudoparticles with repulsive interaction
\cite{Pechukas:1983,Gaspard:1990}.

\begin{figure}[t]
\includegraphics[scale=0.55]{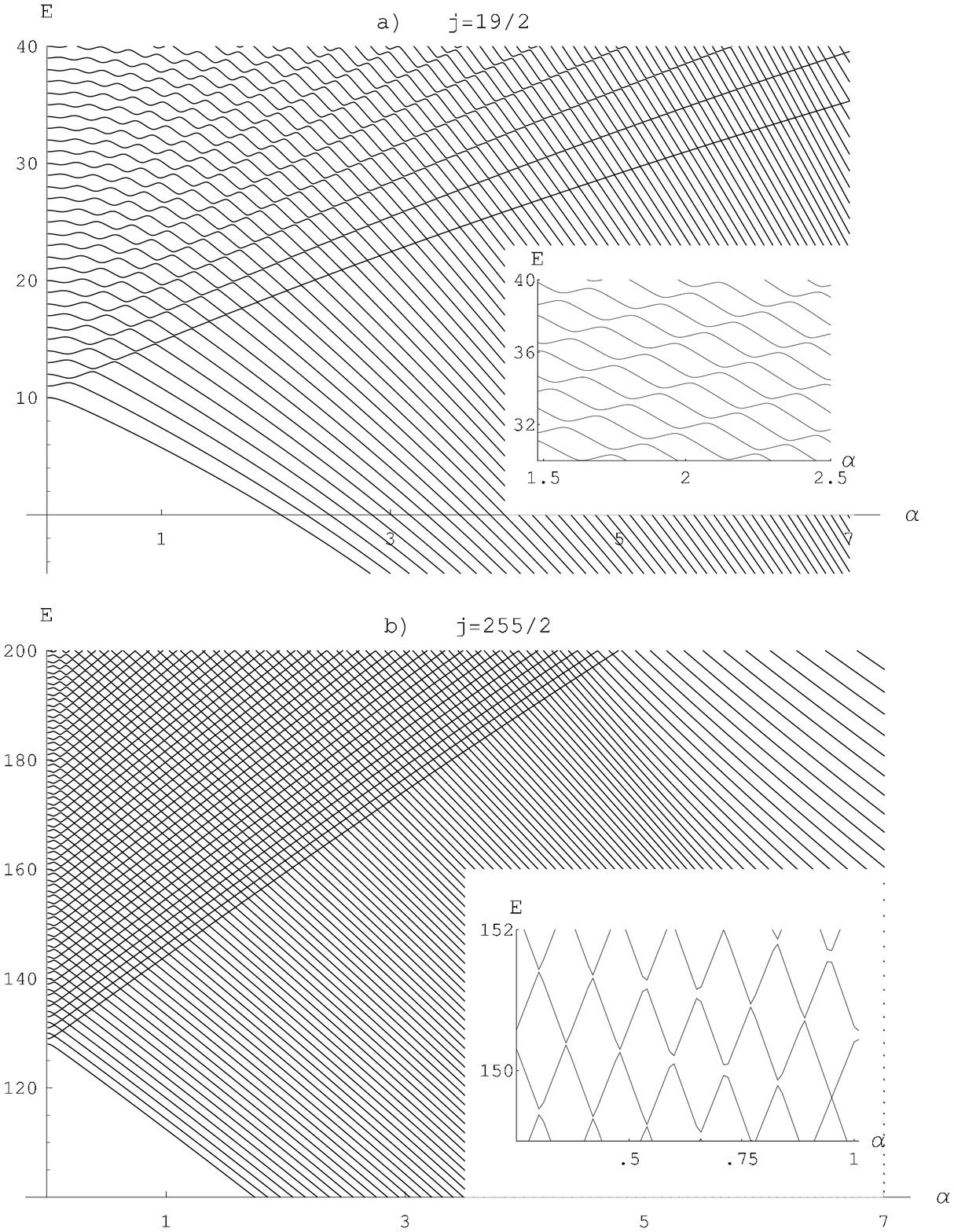}
\caption{Energy spectra for $j=19/2,255/2$ as a function of
$\alpha$.} \label{figS1-b}
\end{figure}

In Sec. II we briefly present an analytical formulation of the
 eigenvalue problem in radial coordinates. Since the literature on one-phonon
models
 is more abundant and certain similarities between both models appear,
  it is useful to point out the essential differences based on a
 comparison of the relevant
  integrable models. In the representative one-boson mode model
considered by Cibils {\it et al.}
  \cite{Cibils:1995} the integrable case was labeled by a pair of
quantum numbers $\mu,
 \nu$ where $\mu=-s,-s+1,\dots, +s$ and $\nu=0,1,2\dots$. In our model
 the integrable case is labeled by three quantum numbers $n_r,j,p$,
 where  $n_r$ is the radial quantum number $n_{r}=0,1,2,\dots$, $j$ is
the rotation quantum number
 $j=1/2, 3/2, 5/2, \dots $
 and $p$ is parity, $p=\pm 1$ for gerade and ungerade states. Thus,
there are two groups
 of states which represent a doubly degenerate set characterized by
parity $p=\pm 1$. Evidently, $n_r$ is analogous to $\nu$, $p$ to
$s=\pm 1/2$, and $j$ has no analog because our oscillator is two
dimensional. The Hamiltonian analysis is supported by numerical
evaluations of the excited energy spectra and corresponding phonon
wave functions in a representation with definite parity and square
of angular momentum. We choose a representation of the spectra and
their probabilistic characteristics (Sec. IV) for definite $j$
because it contains well-defined physical information. However,
the real spectra and probabilistic distributions are composed of
the contributions of all $j$'s.

 In Sec. III we present
approximate analytical approaches to the E$\otimes $e JT model for
the strong and weak intercluster coupling in order to elucidate
the peculiar features of the numerical excited spectra and wave
functions. In particular, we find the approximate analytical form
of wave functions for strongly localized ("exotic") states as well
as for extended states and compare them with respective numerical
results. Special interest is given to identification of the
regions of the spectra with different extent of ordering, from the
long-range ordered region of linear oscillations (extended states)
to the short-range ordered region of localized states with
dynamical avoiding (kink lattice). The dynamical Calogero-Moser
approach appears to be capable of describing all three regions of
the spectra: the dimerization region with pairs of oscillating
levels, the kink train lattice, and  the intermediate region of
nonlinear fluctuations accounting for the kink nucleation
 phase. The intermediate region was found to exhibit quantum chaotic
behavior under certain conditions.
 Section IV complements the dynamical insight by that based on
statistical properties of the spectra. We consider the nearest
level-spacing probability distributions and probability
distributions of level curvatures used as relevant statistical
characteristics of complex spectra
\cite{Eckhardt:1988,Nakamura:1993,Nakamura:1985}. The particular
level-spacing distributions for fixed rotation quantum numbers are
believed to be the most informative, since they behave very
specifically for different parts of the spectra, reflecting their
specific dynamic behavior.

\section{Hamiltonian and equations for excited spectrum}

  We investigate local spinless double degenerate electron states
  linearly coupled to two intramolecular phonon (vibron) modes
  described by Hamiltonian
\begin{equation}
H=  (b_{1}^{\dag}b_{1} +b_{2}^{\dag}b_{2}+1 )I + \alpha
(b_{1}^{\dag}+b_{1})\sigma_{z}
 -\beta (b_{2}^{\dag}+b_{2})\sigma_{x},
 \label{1}
\end{equation}
where $\sigma_x=\left (\matrix {0,\ 1\cr 1, \ 0}\right )$,
$\sigma_y=i\left (\matrix {0, -1   \cr 1,  0}\right )$,
$\sigma_z=\left (\matrix {1, 0 \cr 0,  -1}\right )$ are Pauli
matrices, $I$ is the unit matrix. This pseudospin notation refers
to two-level one-electron (spinless) system.

For general $\alpha\neq\beta$ the Hamiltonian possesses a special
$SU(2)$ symmetry with the reflection operator $\hat{R}$ acting on
phonon and electron subspace:

\begin{equation}
\hat{R}=R_{el}R_{ph}, \quad R_{el}=\sigma_x \,, \quad
R_{ph}=\exp(i\pi b_1^{\dag} b_1) \label{R}
\end{equation}
and $R_{ph}Q_1=-Q_1 R_{ph}$, $R_{ph}Q_2=Q_2 R_{ph}$; i.e.,
phonon $1$ is antisymmetric and  phonon $2$ symmetric against
the reflection. Then, $[R,H]=0$ and the eigenstates can be cast as
eigenfunctions of the reflection operator with eigenvalues
$\pm 1$ which are good quantum numbers.

If $\alpha=\beta$, the Hamiltonian
 becomes rotationally symmetric in the plane $(Q_1\times Q_2)$
and there arises another good quantum number, the eigenvalue
of angular momentum $\hat{J}$ which in fact is nothing but the
infinitesimal generator of the rotational group:

\begin{equation}
\hat{J}=i(b_1 b_2^+-b_1^+b_2)-\frac{1}{2}\sigma_y \,.
\label{J:rot}
\end{equation}
Therefore, in the vicinity of the symmetric JT case it is feasible
to cast the solutions in the form respecting this rotational
symmetry. To account accurately  for the group properties of the
Hamiltonian one introduces radial coordinates in the $Q_1\times Q_2$
plane $Q_1=r\cos\phi\, , \quad Q_2=r\sin\phi.$ The angular momentum
operator (\ref{J:rot}) is
$$
\hat{J}=-i\frac{\partial}{\partial \phi}-\frac{1}{2}\sigma_y \,,
$$
and the phonon part $R_{ph}$ of the reflection operator (\ref{R})
acts as $R_{ph}(r,\phi)f(r,\phi)=f(r,\pi-\phi)$ on some
$f(r,\phi)$ affecting only the $\phi$ coordinate. The eigenfunctions
of the angular momentum operator with eigenvalue $j$ (i.e.,
$\hat{J}\Psi=j\Psi$) can be set as follows:

\begin{equation}
\Psi_j=f_1(r)\cdot |+\rangle e^{i\phi\left(j-\frac{1}{2}\right)}+
f_2(r)\cdot |-\rangle e^{i\phi\left(j+\frac{1}{2}\right)},
\end{equation}
with arbitrary $f_1(r), f_2(r)$ to be determined below.
Electronic wavefunctions are expressed as
$$
|+\rangle=\frac{1}{2}\left(\matrix{1+i \cr 1-i} \right)\,,\quad
|-\rangle=\frac{1}{2}\left(\matrix{1-i \cr 1+i} \right).$$
The
angular quantum number $j$ takes on the values $j=\pm1/2,\pm 3/2,
\pm 5/2, \dots $. The operator $\hat{J}$ does not commute with
the reflection operator $\hat{R}$, but $\hat{J^2}$ does, so it is
more convenient to build solutions as joint eigenfunctions of
$\hat{R}$ and $\hat{J^2}$ rather than $\hat{J}$.

In order to make use of the reflection property of the Hamiltonian we
perform the custom Fulton-Gouterman transformation \cite{Shore:1973} 
by means of
the operator $\hat{U}$:
\begin{equation}
\hat{U}=\frac{1}{\sqrt{2}}\left (\matrix {1,  R_{ph}\cr 1,
-R_{ph}}\right ).
\end{equation}
The resulting Hamiltonian is diagonal in electronic space and is
represented in its full form as follows:

\begin{eqnarray}
\tilde{H}\equiv \hat{U}\hat{H}\hat{U}^{-1}= H_r -\frac{1}{2
r^2}\cdot \frac{\partial^2}{\partial \phi^2}\nonumber\\
+\sqrt{2}\alpha r \cdot \left( \matrix{\cos\phi-\sin\phi R_{ph}  & 0
\cr 0 & \cos\phi+\sin \phi R_{ph} } \right)
\nonumber \\
+ (\alpha-\beta)\sqrt{2}r\cdot \left(\matrix{\sin\phi R_{ph} \ & 0 \cr 0 \ &
-\sin\phi R_{ph}}   \right) \, ,
\label{Ham}
\end{eqnarray}
where
\begin{equation} H_r= -\frac{1}{2r}\frac{\partial}{\partial r}
\left( r\frac{\partial}{\partial r}\right) +\frac{1}{2}r^2
\label{HR}
\end{equation}
is the radial part and $R_{ph}$, Eq.(\ref{R})
 is the (highly nonlinear) phonon
reflection operator in the radial coordinates. In Eq. (\ref{Ham})
we extracted explicitly the last term $\sim (\alpha-\beta)$ which
breaks the rotational symmetry of the problem. Just for reference
we mention the corresponding transformation of the angular
momentum and its square:

\begin{equation}
\tilde{J}=-i\frac{\partial}{\partial
\phi}\sigma_x+\frac{1}{2}\sigma_y R_{ph} \, .
\end{equation}
\begin{equation}
\tilde{J^2}=\left( -\frac{\partial^2}{\partial \phi^2}+\frac{1}{4}
\right) +\frac{\partial}{\partial \phi} \sigma_z R_{ph} . \label{J}
\end{equation}
The transformed electron reflection operator is of course, trivial
{\it per constructionem}, $\tilde{R_{el}}=\sigma_z. $

Although the E$\otimes $e JT model ($\alpha=\beta$) was treated
for the first time by Longuet-Higgins {\it et al.} \cite{Long:1958} a long
time ago, we present briefly the solution using another
representation of the states. This representation enables us to
distinguish the extended and "exotic" (localized) states, as we
explain in Sec.III. The case $\alpha\neq \beta $ can be
included on equal footing. The wave functions we are concerning
with have definite parity $\pm 1$ and square of angular momentum
$j^2$. Let us denote $K=\exp\left(i\pi(j-1/2)\right)=\pm 1.$

i) In the case $K=+1$ ($j-1/2=0,2,4,6,\dots; -2,-4,\dots$) the
{\it gerade} solutions read as

\begin{eqnarray}
\tilde{\Psi}=\left(\matrix{1 \cr 0} \right) \sum_{i=1,2}
\frac{f_i(r)}{2\pi}\left[\sin\left(j+\frac{1}{2}(-1)^i
\right)\phi \right.\nonumber\\
\left.- \cos\left(j+\frac{1}{2}(-1)^i\right)\phi  \right].
\label{psi}
\end{eqnarray}

We shall refer to the angular wave functions in (\ref{psi}) as
$|\pm \rangle\equiv \frac{1}{2\pi}[\sin\left((j+\frac{1}{2}(-1)^i
)\phi \right) \mp \cos\left((j+\frac{1}{2}(-1)^i)\phi\right) ]$.

ii) In the case $K=-1$ ($j-1/2=1,3,5,7,\dots; -1,-3,\dots$) we
have the expression similar to (\ref{psi}) but with the $+$ sign
in the square bracket.

  The {\it ungerade} case can be recovered if one takes the vector
$\left(\matrix{0 \cr 1} \right)$ instead of $\left(\matrix{1 \cr 0}
\right)$. We limit ourselves to the case of {\it gerade} parity
that is, in the representation (\ref{Ham}) we are to take the upper
row of the matrices.

The secular equations for the vibronic part (functions $f_1(r),
f_2(r)$) in both cases are

\begin{equation}
 H_r f_i+\frac{1}{2 r^2}\left(j+\frac{1}{2}(-1)^i \right)^2 f_i+\alpha\sqrt{2}r
f_k=E f_i , \quad i\neq k, \ i, k=1,2 \ ,
\end{equation}
where the radial part $H_r$ is given by (\ref{HR}).

In the case $\alpha=\beta=0$ the problem is exactly that of two
independent oscillators and its solution in terms of special
functions is well known. Thus, the solutions are grouped according
to the $|j|$ value ($j=1/2,3/2,5/2,\dots$; the parity is +1, as
before). Within the same $j$ the functions are divided into two
groups,

(a) States with  $f_2=0$ and
\begin{eqnarray}
f_1 & \equiv & \Phi^{+}(n_r,j;r) \nonumber \\
& =& \frac{1}{\sqrt{C^+_{n_r,j}}} \exp(-\frac{r^2}{2})\cdot
r^{\left| j-1/2 \right|} \cdot _1
F_1\left(-n_r,1+\left|j-\frac{1}{2}\right|,r^2 \right)  \, , \nonumber \\
& & \label{16}
\end{eqnarray}
where $n_r$ ("radial" quantum number) takes on positive integer values
$n_r=0,1,2,\dots$; $_1F_1(a,b,z)$ is a confluent hypergeometric
function [satisfying equation $zy''+(b-z)y'-ay=0$], and the norm is

\begin{equation}
\frac{1}{C^+_{n_r,j}}=\frac{\left(\left|j-\frac{1}{2}\right|\right)!}{2}\cdot
\frac{1}{C^{n_r+|j-1/2|}_{|j-1/2|}} \,
\end{equation}
(here $C^n_m\equiv n!/[m!(n-m)!]$ is a common combinatorial factor).
 We label these states by $+$  as they correspond to the
$|+\rangle$-function of the angular momentum. The product of two
radial functions is defined  as $\int\limits_0^{\infty} r
\Phi^{±}(n_r,j,r)\Phi^{±}(n_r,j,r) \mathrm{d}r.$ The energy spectrum is
$E^+=2n_r+|j-1/2|+1$, and the "main" quantum number is therefore
$n=2n_r+\left|j-1/2\right|$.

(b) States with  $f_1$=0 and $f_2\equiv \Phi^{(-)}(n_r,j;r)$ and
energy $E^-=2n_r+\left|j+\frac{1}{2}\right|+1$ are given by the same
expression as (\ref{16}) but with $j+1/2$ instead of $j-1/2$. We
refer to these  states as $|-\rangle$ states.

The two groups mentioned correspond to two initially degenerate
electron levels; these states are coupled together by means of the
strongly nonlinear transformation involving the reflection
operator $R(r,\phi)$; thus, the situation is similar to the
Foulton-Gouterman treatment in ordinary phonon coordinates, where
the nonlinear reflection operator coupled the phonon states
pertaining to the lower and upper electron level
\cite{Majernikova:2003}.

Writing secular equations for the problem with $\alpha\neq 0$ is
straightforward. The solution is sought as $f_1=\sum_{n_r}
a_{n_r}\Phi^+_{n_r}$, $f_2=\sum_{n_r} b_{n_r} \Phi^-_{n_r}$, or, in
unified notation, $f=\sum\limits_{n=0}^{\infty}c_n\Phi_n,$ where
$\{c_n\} \equiv \{a_0,b_0,a_1,b_1,a_2,b_2,a_3,\dots\}$, and
\begin{equation}
\Phi_n = \left\{\matrix {\Phi^+(n_r,j,r), \quad n=2n_r,\cr
\Phi^-(n_r,j,r), \quad  n=2n_r+1. }\right.
\end{equation}

The only non-zero off-diagonal elements are

\begin{eqnarray}
f_{n_r n_r}\equiv\sqrt{2}\int\Phi^+_{n_r}\Phi^-_{n_r} r\cdot r \mathrm{d}r=
\sqrt{2}\sqrt{n_r+1+ \left| j-\frac{1}{2} \right|},
\label{f:nn}\\
f_{n_{r}+1 n_r}\equiv\sqrt{2}\int\Phi^+_{n_r+1}\Phi^-_{n_r} r\cdot r
\mathrm{d}r=-\sqrt{2}\sqrt{n_r+1}. \label{f:nnn}
\end{eqnarray}

Therefore, the secular equation  acquires the familiar tridiagonal form
($E^{(0)}$ is "unperturbed" energy for the corresponding $n,j$),
\begin{eqnarray}
c_nE^{(0)}_n+\alpha(c_{n+1}f_{n_r n_r}+c_{n-1}f_{n_r n_r-1}) =E
c_n, \quad  n=2n_r, \label{27}\\
c_nE^{(0)}_n+\alpha(c_{n-1}f_{n_r n_r}+c_{n+1}f_{n_r+1 n_r}) =E c_n,
\quad n=2n_r+1
 \label{28}
\end{eqnarray}
with off-diagonal coefficients given by Eqs. (\ref{f:nn}) and (\ref{f:nnn}).
The asymmetry of the coefficients with increasing $j$ is crucial for
the appearance of nonlinear effects in spectral properties of the model.
 We remark by passing that Eqs.  (\ref{f:nn}) and (\ref{f:nnn}) 
differ from those given in
\cite{Long:1958} by the ``minus'' sign near  $f_{n+1 n}$. Changing
this sign to plus means merely redefinition of the base wave
functions (every even base function acquires the opposite sign)
and does not affect the energy spectrum. In the following it will
be more convenient to use the $+$ sign of both expressions
(\ref{f:nn}) and (\ref{f:nnn}).

\begin{figure*}[t]
\includegraphics[scale=0.9]{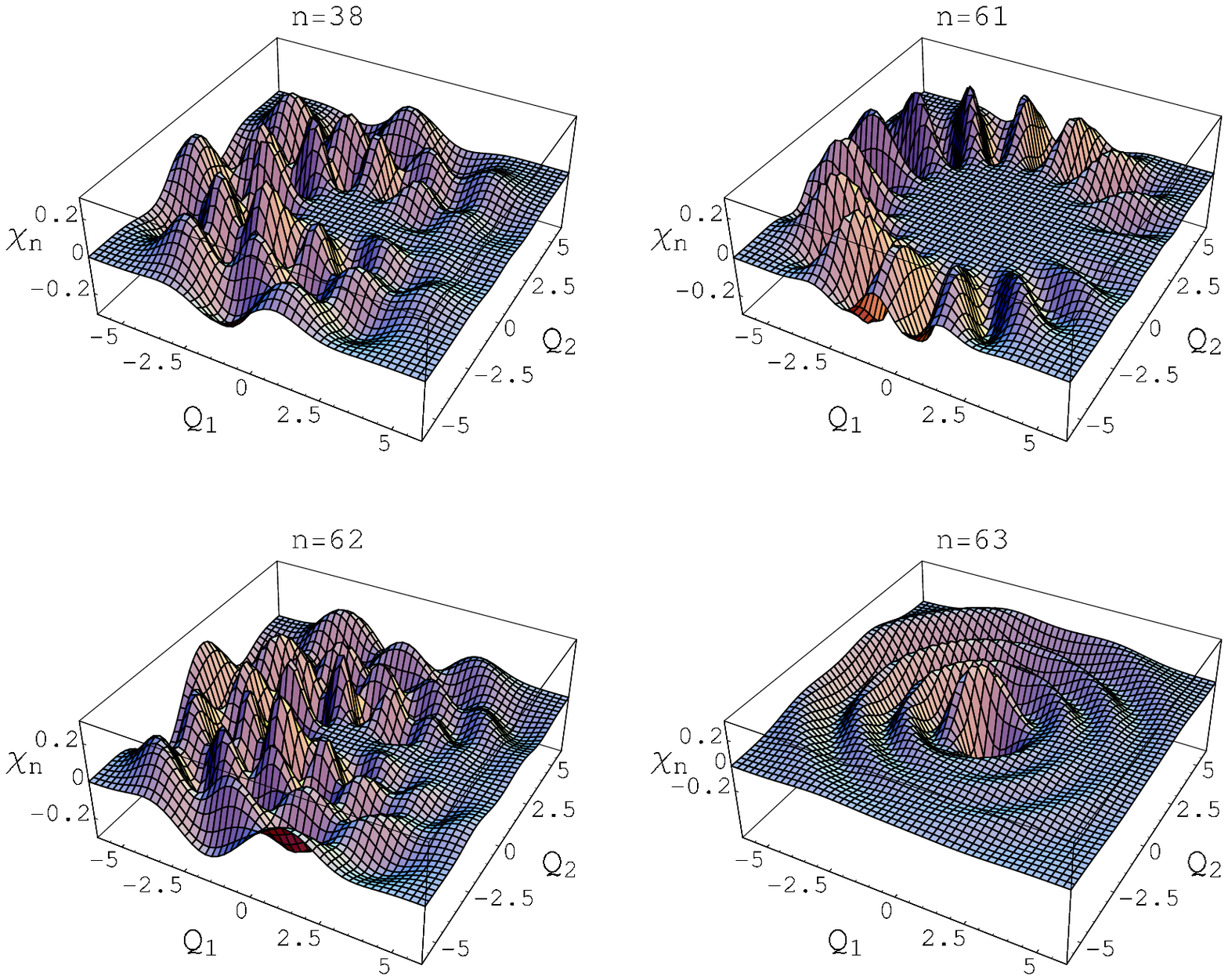}
\caption{(Color online) Numerical wave functions $\chi_n$ in the
plane $Q_1\times Q_2$, $\alpha=\beta=2$. States n=61 and 63 are
localized ("exotic"). } \label{figW1-d}
\end{figure*}

By  numerical diagonalization of the system (\ref{27}) and (\ref{28}) we
obtain the phonon excited energy spectrum  as function of $\alpha$
(we have taken $n$ up to $\sim 1200$), as well as the corresponding
wave functions. Figure 1 shows examples of the spectrum as
function of $\alpha$ for $j=19/2,255/2$). The sample numerical
wave functions $\chi_n$ of a few excited states of (\ref{1}) are
shown in Figs. 2 and 3. Figure 2 presents the wave functions $\chi_n
$ in phonon coordinate representation (in the plane $Q_1\times
Q_2$), while Fig. 3 shows the projection $\chi_{n,m}\equiv
|\langle\chi_n|\Phi_m^0\rangle|^2$ of the exact numerical wave
functions on the unperturbed ($\alpha=0$) states of the radial
oscillator (\ref{16}) \cite{ftn}. The right-hand side of Fig. 4
visualizes the whole picture of wavefunctions at a given $\alpha,j$
by assigning to each state $\vec{C}^{(n)}\equiv \{C_1, C_2,\dots ,
C_m, \dots \}$ the spectral entropy $S=-\sum_m C_m^2\log C_m^2$
(which gives the width of the wavefunction in the representation
of Fig. 3).

\begin{figure*}[bht]
\includegraphics[scale=0.7]{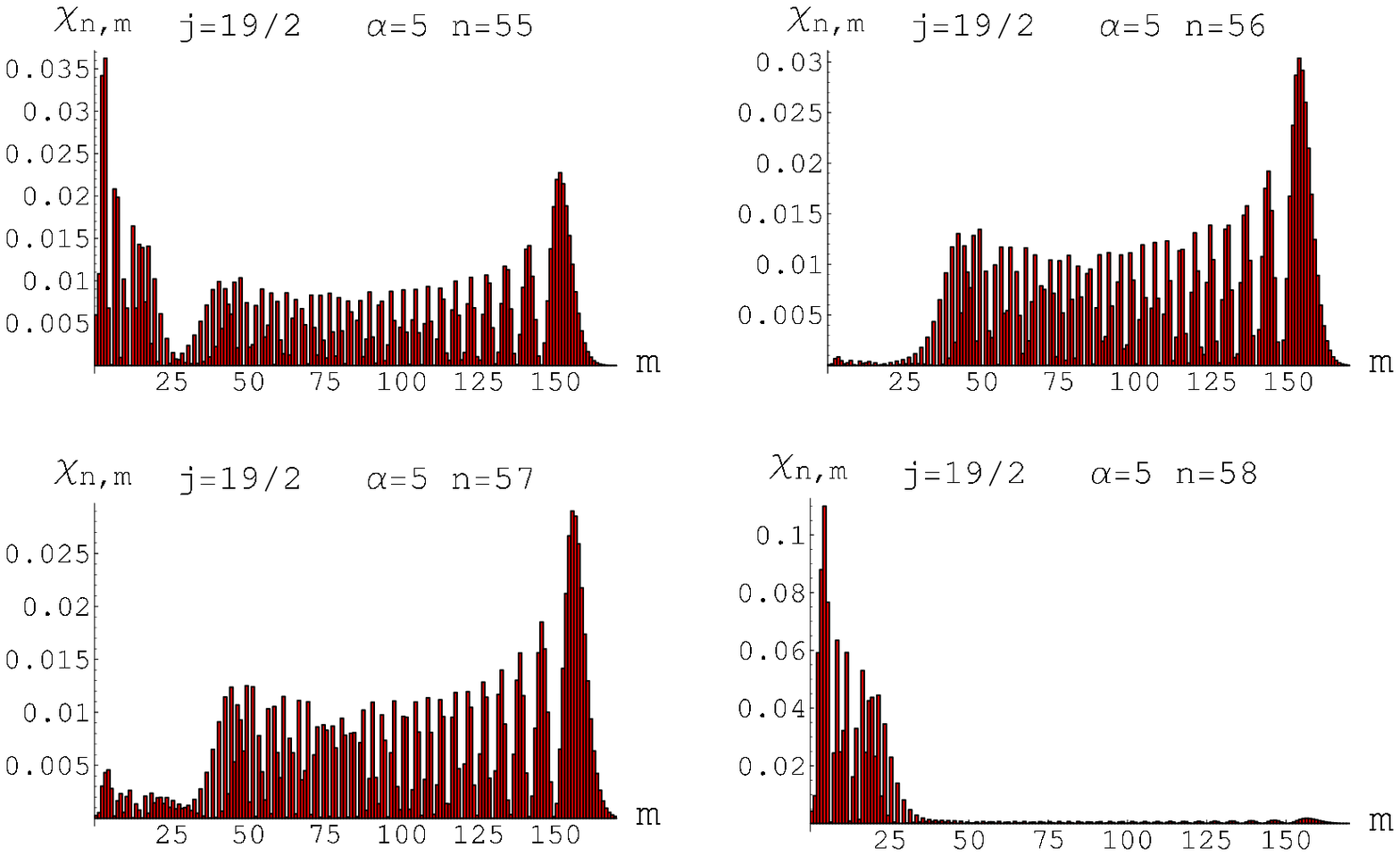}
\caption{(Color online) Projections of exact symmetric
($\beta=\alpha$) states $\chi_n$ on the harmonic oscillator base,
$\chi_{n,m}=|<\chi_n|\Phi^0_m>|^2$. } \label{figW2-b}
\end{figure*}

One remarkable feature of excited states in JT models (both
rotation symmetric and that with  broken rotation symmetry) are
strongly localized (``exotic'' \cite{Wagner:1992}) states emerging
abruptly between more extended ones. These states are exemplified
in Fig. 2 ($n=61, 63$) and in Fig. 3 ($n=58$ for $j=19/2$). The
localization of those states is remarkable in comparison with 
neighboring
or even lower-lying ones (like $n=38, 62$ from Fig.2 compared to
$n=61,63$).

\begin{figure}[h]
\includegraphics[scale=0.55]{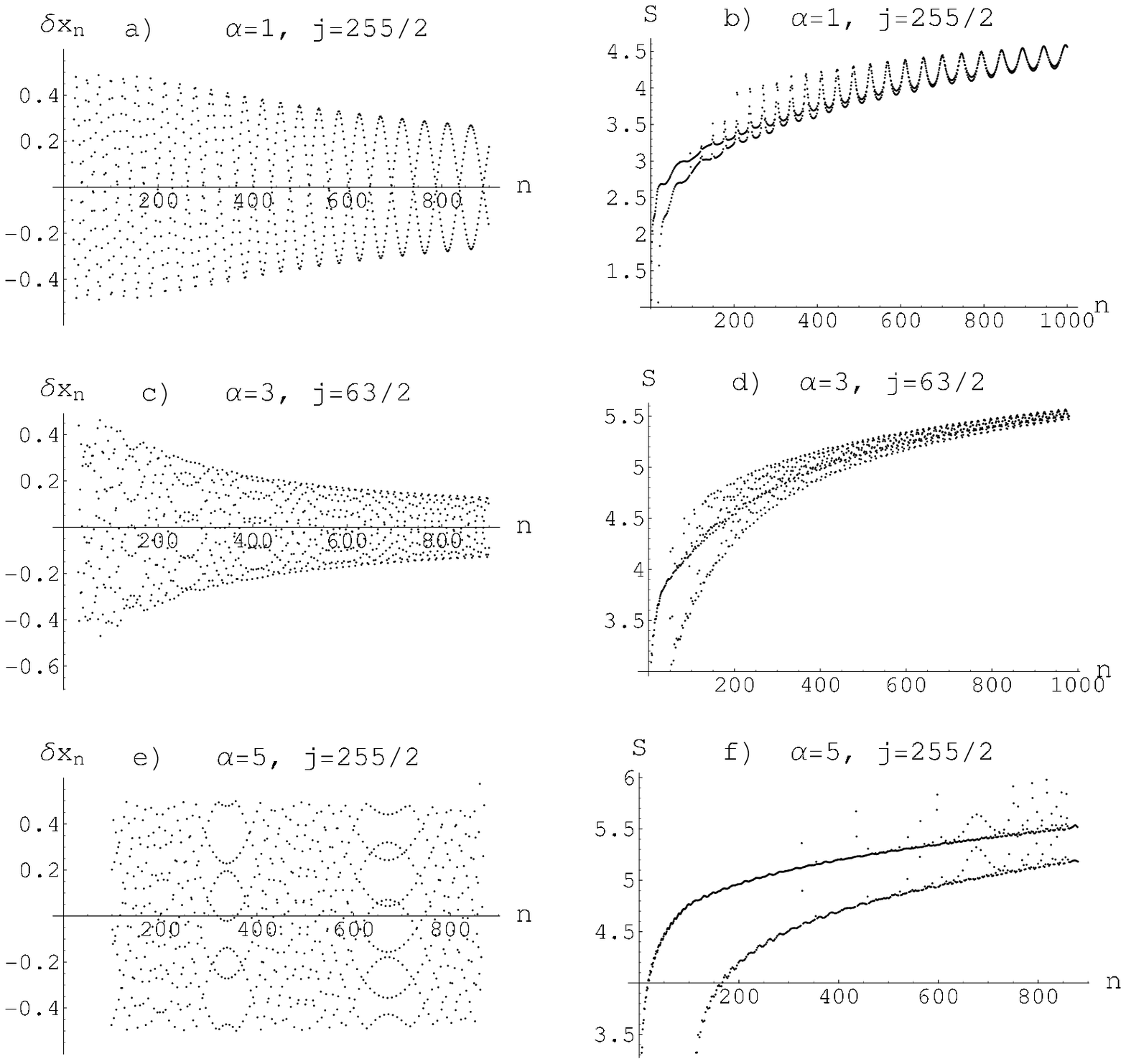}
\caption{Reduced  energies (see text) and spectral entropies of
wave functions $\chi_n$ for different couplings $\alpha $ and $j$.
} \label{figW5}
\end{figure}

There is an intimate relation between the position of localized states
and spectral picture of levels (Fig.1). Closer inspection [see the
insets of Figs. 1(a),1(b)] shows that each and every energy level taken
for a given $j$ is avoided. Varying the parameter $\alpha$
infinitesimally slowly will cause the system to follow one and the
same adiabatic level. But on larger scale the spectra of Fig.1
show two sets of "diabatic" lines \cite{Nakamura:1993} (directed
upwards and downwards) which reflect a different behavior of the two main
wells of the effective potential \cite{Majernikova:2003} when
changing $\alpha$. From the right-handside of Fig.4 it is 
possible to note that the
states for large $\alpha,j$ are grouped into two separated branches
and the domain of localized states with small spectral entropy
emerges at a critical energy corresponding to the appearance of
the upper sheet of the effective potential. If one varies the
parameter $\alpha$ fast enough the system will follow the
"diabatic" lines changing the energy level number so that its wave
function remains almost untouched which means that the system
resides on the corresponding sheet of effective potential. All
"exotic" states appear to pertain to the directed upwards diabatic
lines of the spectrum seen markedly in Fig. 1. The multitude of
avoided crossings (and therefore the domain referred to as
"quantum chaotic") is thus formed in the intermediate energy
region where these two ``phases'' coexist-- that is above the first
(lowest) diabatic line. It is clearly seen in Fig. 1(b) for large
values of the angular momentum number $j$, while for small values of
$j$ these exotic states, although still existing, do not come into
complex interplay with the ordered states. This picture is in
agreement with the semiclassical treatment of the problem where
the classically chaotic region is known to emerge sharply above
some critical energy.

\section{Analytical quantum treatment}

\subsection{Strong-coupling approximation: ``Exotic'' states}

The peculiarities of the spectra shown in Fig. 1 as well as their
connection to the character of pertaining wave functions (Figs. 2
and 3) can be understood from merely a rough approximation to the
set of equations (\ref{27}) and (\ref{28}) in the following way. From
Eqs. (\ref{f:nn}) and (\ref{f:nnn}), for $|j| \gg n_r$, there holds
$|f_{n_r,n_r}|\gg |f_{n_r+1,n_r}|$, and the basis functions in
Eqs.(\ref{27}) and (\ref{28}) appear to be ``clustered'' in the sense that
the coefficients $C_{2 n_r} and C_{2 n_r+1} $ pertaining to the same
$n_r$ are coupled together (through $f_{n_r,n_r}$) essentially
more strongly than to neighboring elements. [In order to avoid
confusion we note that the terms "strong-" and "weak-coupling-approximation" 
for this section are not meant in the usual sense of
large and small values of the parameter $\alpha$ but refer to the
relative coupling inside and between the "clusters"
$(C_{2n},C_{2n+1})$). They are determined by the relation between
$f_{nn}$ and $f_{n+1 n}$-- i.e., by the relative values of quantum
numbers $j$ and $n$]. Therefore, for the zeroth approximation of
strong coupling $f_{n+1 n}=0$, and the reduced version of
Eqs. (\ref{27}) and (\ref{28}) for each $n_r$ gives us two sets of localized
solutions for the energy
\begin{equation}
E_{n_r,j}(\alpha) = 2 n_r+|j-1/2|+3/2\pm \sqrt{2 (n_r+1+|j-1/2|)}
\alpha \, , \qquad n_r=0,1,\dots \label{E:diabatic}
\end{equation}
with corresponding localized eigenvectors $C_{2n_r+1}\simeq
C_{2n_r}$ ($+$ sign) and $C_{2n_r+1}\simeq -C_{2n_r}$ ($-$ sign) ,
the other $C_n$ being zero. From Fig. 1 we can conclude that the slopes of
the corresponding lines follow the predictions given by
Eq.(\ref{E:diabatic}). This is valid for several lower clusters with
small $n_r$ which show up as directed upwards and downwards "diabatic"
lines. At higher $n_r$, when $j \leq n_r$ the intercluster
$f_{n_r+1 n_r}$ and inside-cluster elements $|f_{n_rn_r}|$ become
comparable and the long-range correlations yield the extended
wavelike solutions seen in the upper parts of spectra in Figs.
1(a) and 1(b) [cf. inset of Fig.1(a)].

To find the form of localized wavefunctions along the diabatic
line we proceed by treating terms with $f_{n+1 n}$ as perturbation
to the zeroth-order solution (\ref{E:diabatic}). This approximation
is valid far from the points of avoided crossing where the
wave functions are essentially localized and unaffected by
neighbors. Let us denote $\delta E \equiv E - E_m^*$ where $E_m^*$
is the zeroth-order strong-coupling solution (\ref{E:diabatic}) for
a given $m$ and use the strong-coupling ansatz $C_0=C_1, \dots ,
C_{2n}=C_{2n+1}$. The sum of Eqs.(\ref{27}) and (\ref{28}) then yields

\begin{equation}
[2(n-m) + \alpha (f_{nn}- f_{mm})] C_{2n} +
\frac{\sqrt{2}}{2}\alpha \left(\sqrt{n+1}C_{2n+2} +
\sqrt{n}C_{2n-1} \right) = \delta E C_{2n} \,.
\label{strong:couple:2}
\end{equation}

Further, let us {\it formally} mark the cluster $(C_{2n},
C_{2n+1})$ as $|n\rangle$ and introduce formal creation and
annihilation boson operators $\hat{b}^+$ and $\hat{b}$ acting on these
clusters. Since $j \gg n$, the term $f_{nn} - f_{mm}$ can be
simplified by keeping  only the leading-order term $\sim (n-m)$. Then
the effective Hamiltonian equation in strong-coupling
approximation is ($n |n\rangle$ is replaced by $b^+b |n\rangle$)

\begin{equation}
\left[ \Omega^{(+)}_{eff} b^+b  + \frac{1}{\sqrt 2}\alpha
\left(\hat{b}^+ + \hat{b} \right) \right] |n\rangle = (\delta
E_n+\Omega^{(+)}_{eff}m)|n \rangle \ , \label{strong:couple:3}
\end{equation}
with
\begin{equation}
\Omega^{(+)}_{eff}\equiv  2 + \alpha/\sqrt{2j-1} \,. \label{Omega}
\end{equation}

 The resulting system presents a quantum oscillator displaced by the term
$\alpha/\sqrt{2}$. The effect of nonlinear fluctuations is
reflected in the {\it squeezing of its effective frequency}
(\ref{Omega}) due to  $j$ down to its minimum $\Omega^{(+)}_{eff}
\rightarrow 2$ in the quantum limit $j \rightarrow \infty$. The
term with $m$ in fact labels the "sites" of a pseudolattice. In
this approximation the solutions for different $m$ are translation
invariant. The subsequent solutions of Eq.(\ref{strong:couple:3}) are
obtained in a standard fashion applying the displacement operator
$D(\gamma)\equiv\exp(\gamma(\hat{b}^+-\hat{b}))$,
$\gamma=-\alpha/(\sqrt{2}\Omega^{(+)}_{eff})$ on the clustered
pseudostates $|n\rangle\equiv (C_{2n}, C_{2n+1})$, giving the
localized functions on subsequent diabatic lines (for example, for
$j=255/2$ and $\alpha=2$ the lowest states of this fictitious
oscillator correspond to wave functions with levels $n=37,39,41,43,
45,48, \dots$; such exact wave function is exemplified in Fig. 3
for $n=58$). We performed an extended comparison of the exact wave functions
to those found via this method in a wide range of parameters $\alpha$
and $j$.
Our calculations show that the solutions of
Eq.(\ref{strong:couple:3}) yield a very plausible approximation to the exact
solutions of Eqs.(\ref{27}) and (\ref{28}) for localized states, especially
for big $j$: for example, for several lowest diabatic states the
projection $\langle
\Phi | \Psi \rangle$ of the approximate over the exact state for
$j=255/2$ lies within $0.999-0.995$. The same method is also
suitable for approximate solutions found for several strong-coupled
extended states, but there the cluster ansatz is antisymmetric,
$C_{2n+1}=-C_{2n}$. The corresponding equation differs from
Eq.(\ref{strong:couple:2}) by the $-$ sign near both terms with
$\alpha$, respectively; there appear the $-$ sign of $\alpha$ in
Eq.(\ref{strong:couple:3}) and in the definition of
$\Omega^{(-)}_{eff}$
\begin{equation}
\Omega^{(-)}_{eff}\equiv  2 - \alpha / \sqrt{2j-1} \, ,
\end{equation}
 which is now {\it antisqueezed} due to nonlinear fluctuations.
In the limit $j\rightarrow\infty $ both oscillators degenerate at
 frequency $2$ and so the intercluster interaction effectively
vanishes. In the absence of the intercluster interaction the energy
lines from different clusters would regularly cross. The
intercluster interaction leads to the spreading of the localized
wavefunctions as described above and to the splitting of levels
(avoided crossings). In the vicinity of an avoiding crossing we can
approximately represent the relevant Hamiltonian as an effective
($2\times 2$) matrix:
\begin{equation}
\hat{H}= \left(
\begin{array}{ll}
p \tau & u \\ u & 0
\end{array}
\right) \ ,\label{sol:matr}
\end{equation}
where $\tau=\alpha-\alpha_0$, $\alpha_0$ is the point of level
collision if there were no interaction.  The matrix
(\ref{sol:matr}) accounts for the collision of two levels, one
lying at $E=0$ (if $\tau=\pm \infty$) and the other moving
asymptotically with velocity $p$ with "pseudotime" $\tau$. In the
representation (\ref{sol:matr}) we are in the frame of reference
moving downwards with $\alpha$ so that the velocity of the
nonexotic levels is zero. If the colliding levels are weakly
affected by neighbors we can consider $p$ and $u$ to be
approximately constants. The minimal avoided crossing at $\tau=0$
is $2|u|$. Including other non-exotic levels (thus situated
at distances $1,2,\dots$) leads to the extension of the matrix
(\ref{sol:matr}) to the shape given by Eq. (2) of the paper
by Gaspard {\it et al.} \cite{Gaspard:89} and accounting for a single
level moving with velocity p and crossing subsequently parallel
energy levels $y_1, y_2, \dots$ with effective interaction
strengths $u_1, u_2, \dots$.
 It was shown there that for equidistant energies $y_i$ and
equal strengths $u_i$ this system immediately gives a solution in
the form of a soliton propagating with pseudotime $\tau$ through the
``lattice'' $y_n$ with natural analytical predictions as to the
profile of energy eigenvalues $x_n(\tau)$. This soliton like
solution corresponds to our diabatic level. The parameter $p$ equals
the relative slope of exotic (diabatic) and ordinary level and
can be estimated in the strong-coupling approximation using
Eq.(\ref{E:diabatic}); for the first diabatic line, it therefore holds
$p \simeq 2\sqrt{2j}.$ In the same approximation the level splitting
(parameter $u$) can be estimated as $\langle \Psi_n |
\hat{H}(\alpha_0) | \Phi_m \rangle$, where $\hat{H}$ is the exact
Hamiltonian and $\Psi_n$, $\Phi_m$ are wave functions in the
strong-coupling approximation of the localized and extended states
with numbers $n$ and $m$, $n=1,2 \dots$, $m=n+1,\dots$ taken in the
point $\alpha_0$ of crossing of the corresponding unperturbed levels.
For example, for the first avoided crossing of the lowest diabatic
line the estimation gives $u \sim 2/\sqrt{j}$. For large numbers $m$
of subsequent collisions the gap value turns out to be approximately
constant which suggests the applicability of the soliton like
solution like those considered by Gaspard {\it et al.} \cite{Gaspard:89}.

\subsection{Weak-coupling approximation}

This is the case opposite to the previous one and valid for $j \ll
n_r$. Rewrite the system (\ref{27}) and (\ref{28}) with coefficients
(\ref{f:nn}) and (\ref{f:nnn}) with positive signs in the following way
(here we again label by $n$ instead of $n_r$):

\begin{equation}
\left\{
\begin{array}{l}
C_{2n}(E_{2n}^0-E) + \alpha \left[ C_{2n+1}\sqrt{2n+1} +
C_{2n-1}\sqrt{2n} \right] + \alpha C_{2n+1} k_n
 =  0, \\
C_{2n+1}(E_{2n+1}^0-E) + \alpha \left[ C_{2n}\sqrt{2n+1} +
C_{2n+2}\sqrt{2n+2} \right] + \alpha C_{2n} k_n =  0
\end{array}
\right.
 \label{weak:eq1}
\end{equation}
with
\begin{equation}
k_n \equiv \sqrt{2n+1} \left( \sqrt{1+\frac{2j}{2n+1}}-1\right)
  \ ,\  n=0,1,\dots
\label{weak:kw}
\end{equation}
$\left(k_n \sim j/\sqrt{2n+1} \ll 1 \right.$ if $j\ll n$). The
whole system (\ref{weak:eq1}) is elegantly cast as $\hat{H} \Psi =
E \Psi$ with

\begin{equation}
\hat{H}= \hat{b}^+\hat{b}+\alpha\left(\hat{b}^+ + \hat{b} \right)
+ \delta \hat{V} \equiv \hat{H}_0 + \delta \hat{V} \,
\label{weak:eq2}
\end{equation}
with {\it formal} notations: $C_{2n} \to |2n \rangle$ and 
$C_{2n+1} \to |2n+1\rangle$ and operators $\hat{b}^+, \hat{b}$
acting upon our radial functions  $|n\rangle$ as ordinary creation
and annihilation operators [in the previous subsection the
corresponding "pseudostates" referred to the cluster
$(C_{2n},C_{2n+1})$]; the perturbation operator $\delta \hat{V}$
acts according to the  recipe:
\begin{equation}
\delta \hat{V} \left\{
\begin{array}{l}
|2n\rangle \\
|2n+1 \rangle
\end{array}
\right. = \alpha k_n \cdot \left\{
\begin{array}{l}
|2n+1 \rangle \\
|2n \rangle
\end{array}
\right. \label{weak:pert}
\end{equation}
The interaction (\ref{weak:pert}) ties together the elements
inside the cluster $(C_{2n},C_{2n+1})$ and when it becomes
considerable  we again recover the strong-coupling approximation
of the previous section.

From Eq.(\ref{weak:eq2}) we immediately get the solution of the
unperturbed problem $H_0 \psi = E_0 \psi$ in terms of the
displaced Fock states by acting the coherent operator
$D(\gamma)\equiv \exp(\gamma(b^+ - b))$ on our pseudostates:

\begin{equation}
|\Psi\rangle  \equiv | \tilde{n}\rangle = D(-\alpha) |n>
\label{weak:fun:0}
\end{equation}
with unperturbed energies $ E_{\tilde n}^{(0)}= n - \alpha^2$.
Therefore it is reasonable to investigate the reduced spectrum
from which the "secular parts" of energies are subtracted: $\delta
x_n\equiv E_n - n + \alpha^2 - (j-1/2)$. On the left-hand side of
Fig. 4 we plot the reduced wavelike spectrum $\delta x_n$ as
function of $n$ for typical values of $\alpha, j$. Examining Fig.
4 we observe that extracting the secular part reveals patterns which
have not been noticed on the normal scale of the wavelike
spectrum. The most interesting peculiarity is thus a crossover
from two twisting quasi-sinusoids for large $n$ and/or small
$\alpha$ to the pictures which curiously resemble chaotic
intermittent patterns containing rather characteristic windows of
a regular motion. The cutoffs on the left parts of the spectra
(small $n$) correspond to the first diabatic line below which such
a reduction looses its sense since the perturbation $\delta
\hat{V}$ can no more be considered as small compared to $H_0$.

The standard perturbation scheme applied to (\ref{weak:eq2}) gives
up to the first perturbation order the expression for the spectral
energies the notation $\tilde{n}$ merely reminds one of the fact that the
base functions are now ``displaced Fock states''
(\ref{weak:fun:0}) not to be confused with $|n\rangle$]

\begin{equation}
E_{\tilde{n}}= \tilde{n} - \alpha^2 + \delta
V_{\tilde{n}\tilde{n}} + \dots \label{weak:en:1}
\end{equation}
with
\begin{eqnarray}
\delta V_{\tilde{n}\tilde{n}} = \sum\limits_{h,s} \langle
\tilde{n}|h \rangle \langle h | \delta\hat{V}|s \rangle \langle
s|\tilde{n} \rangle \nonumber \\
= 2\alpha \sum\limits_{h=0}^{\infty}
\langle 2h+1 | D(-\alpha) | n \rangle \cdot \langle 2h | D(-\alpha)
| n \rangle\cdot k_h,  \label{weak:en:2}
\end{eqnarray}
where $k_n$ is defined by (\ref{weak:kw}). The matrix elements
$\langle n | \tilde m \rangle \equiv \langle n | D(\beta) | m
\rangle$ can be calculated directly:
\begin{equation}
 \langle n |D(\beta) | m \rangle
= \exp\left(-\frac{\beta^2}{2}\right) \beta^{|n-m|}\cdot
\frac{sgn(n-m)}{\sqrt{m!n!}}L_{min (n,m)}^{|n-m|}(\beta^2) ,
\label{Dnm}
\end{equation}
where $L_{a}^{b}(x)$ are Laguerre polynomials.

 In the scope of the present paper we do not present further
detailed analysis of Eq.(\ref{weak:en:2}), just mentioning now that it
gives essentially the same structure as that presented in Fig. 4.
Moreover, it can be used even for the ground state (lowest energy
for given $j$) giving a plausible fitting for very small and very
large $\alpha$, although serious discrepancies occur for $\alpha\sim
1$ rendering it absolutely unsatisfactory in this region.

\subsection{Method of level dynamics}

The formalism of the generalized Calogero-Moser gas has been developed
\cite{Pechukas:1983,Gaspard:1990,Nakamura:1993}
 as a useful alternative approach  for Hamiltonians of the type
 $H=H_0+\alpha V$ [the perturbation term linearly dependent on $\alpha$;
 Eq.(\ref{Ham}) for $\alpha=\beta$ is the case with the first two
 terms as $H_0$ and the term with $\alpha$ as $V$]
 by mapping the eigenvalue equations on a
 statistical many-body system of interacting pseudoparticles.
 There $\alpha$ is a controlling
parameter considered as a pseudotime, $\alpha\equiv\tau$. The
eigenvalues were defined as dynamic variables $E_n(\alpha)\equiv
x_n(\tau)$, $dx_n/d\tau= p_n(\tau)$. The respective classical dynamic
equations for the pseudoparticles evolving in the pseudotime
$\tau$ are

\begin{eqnarray}
\frac{dp_n}{d\tau}=2\sum_{m(\neq
n)}\frac{L_{nm}L_{mn}}{(x_m-x_n)^3}\nonumber\\
\frac{dL_{mn}}{d\tau}=\sum_{l\neq
(m,n)}L_{ml}L_{ln}\left[\frac{1}{(x_n-x_l)^2}-\frac{1}{(x_m-x_l)^2}\right
], \label{C1}
\end{eqnarray}
where
\begin{eqnarray}
p_n\equiv  \langle n(\tau)|V|n(\tau)\rangle\equiv V_{nn}\, , \nonumber \\
L_{mn}(\tau) =(x_n(\tau)- x_m(\tau))\cdot V_{mn}=-L_{nm} \,.
\end{eqnarray}

The system of differential equations in the form (\ref{C1}) is
equivalent to the initial set (\ref{27}) and (\ref{28}) and is
represented as a set of dynamical equation of motion for a gas
subject to a repulsive $1/r^2$ potential ("Calogero-Moser gas").
In our case from Eqs.(\ref{f:nn}) and (\ref{f:nnn}) we get (we simplify
the notation, $n\equiv n_r$)
\begin{eqnarray}
L_{2n 2n+1}= \sqrt 2
\sqrt{n+1+|j-1/2|}\cdot(x_{2n+1}(\tau)-x_{2n}(\tau)),
\nonumber \\
L_{2n-1 2n}= -\sqrt 2
\sqrt{n+1}\cdot(x_{2n}(\tau)-x_{2n-1}(\tau))\,. \label{C2}
\end{eqnarray}
The dynamic equations for a pair of even and odd coupled trajectories
are then
\begin{eqnarray}
\frac{dp_{2n}}{d\tau}=2\frac{L_{2n 2n+1}\cdot L_{2n+1
2n}}{(x_{2n+1}-x_{2n})^3}+2\frac {L_{2n 2n-1}\cdot L_{2n-1
2n}}{(x_{2n-1}-x_{2n})^3}
\nonumber\\
\frac{dp_{2n+1}}{d\tau}=2\frac{L_{2n+1 2n+2}\cdot L_{2n+2
2n+1}}{(x_{2n+2}-x_{2n+1})^3}+2\frac {L_{2n+1 2n}\cdot L_{2n
2n+1}}{(x_{2n}-x_{2n+1})^3}\,. \label{C3}
\end{eqnarray}
For what follows we define a small fluctuation $\delta $ by
$x_{2n+1}-x_{2n}\equiv 1+\delta_{2n}$ and approximate
$\left(x_{2n+1}-x_{2n}\right)^{-1}\approx
1-\delta_{2n}+\delta_{2n}^2-\delta_{2n}^3$. From Eqs.(\ref{C2}) and
(\ref{C3}) we get the nonlinear set of equations
\begin{eqnarray}
\frac{\partial^2\delta_{2n}}{\partial
\tau^2}-v_1^2(\delta_{2n+1}+\delta_{2n-1}-2\delta_{2n}) \nonumber \\
=
f-4\left|j-\frac{1}{2}\right|(\delta_{2n+1}+\delta_{2n-1})
-4\delta_{2n-1}\nonumber\\
-4(n+1)(\delta_{2n+1}^2-\delta_{2n+1}^3)-4n(\delta_{2n-1}^2
-\delta_{2n-1}^3)\,, \label{2n}
\end{eqnarray}
\begin{eqnarray}
\frac{\partial^2\delta_{2n+1}}{\partial
\tau^2}-v_2^2(\delta_{2n+2}+\delta_{2n}-2\delta_{2n+1}) \nonumber 
\\
=
-f+4\left|j-\frac{1}{2}\right|(\delta_{2n}+\delta_{2n+2}) \nonumber \\
+4\delta_{2n+2}-4(\delta_{2n+2}^2-
\delta_{2n+2}^3)-8\left|j-1/2\right|(\delta_{2n+1}^2-\delta_{2n+1}^3)\,,
\label{11a}
\end{eqnarray}
where $v_1^2=4(n+1+|j-1/2|)$, $v_2^2=4(n+1)$, and $f=4+8|j-1/2|$.

For large $n\gg |j|$ and $v_1^2\approx v_2^2$,Eqs. (\ref{2n})
and (\ref{11a}) exhibit wavelike character in $n$.  The linear
approximation to (\ref{11a}) bears the solution $\delta_{2n+1}\approx
J_0(2\sqrt{|2j-1|n})$ where $J_0(z)$ is the Bessel function. This
behavior can be recognized in the upper part of the spectra in Figs.
1(a) and 4(a). It is equivalent to the weak-coupling limit of
the preceding subsection. The level avoidings in this linear part
are kinematical and lead to clustering (dimerization) visualized in
the level-spacing probability distributions (Sec.IV).

Equations (\ref{2n}) and (\ref{11a}) allow a useful insight into
the effects of nonlinearities on the level dynamics involved.
Exploiting their wavelike character we define the phase variable
$\zeta= n-v\tau$ ($v$ being an arbitrary constant velocity) and
write an approximate equation for
$\bar\delta_{2n+1}=\delta_{2n+1}-1$ at $|j|\gg n$:
\begin{equation}
-(v^2-4|j|)\frac{\partial^2\bar\delta_{2n+1}}{\partial
\zeta^2}+16|j|(\bar\delta_{2n+1}+\bar\delta_{2n+1}^2)\approx 0 \,.
\end{equation}
 If $4|j|< v^2$, then

\begin{equation}
\bar\delta_{2n+1}(\zeta-\zeta_0) = -\frac{3}{2}\cosh^{-2}\left
(\frac{(4|j|)^{1/2}}{(v^2-4|j|)^{1/2}}(\zeta-\zeta_0)\right ) \ .
\label{fluc}
\end{equation}
There $\zeta_0=n_0-v\tau_0$ is a constant restoring the
translational invariance of the solution;
identifying $\partial x_{2n+1}/\partial n= 1+\delta_{2n+1}$
we recover a soliton-shaped (kink) solution
\begin{eqnarray}
x_{2n+1}=  2n -\frac{3}{2}\frac {(v^2-4|j|)^{1/2}}{(4|j|)^{1/2}}
\tanh
\left(\frac{(4|j|)^{1/2}}{(v^2-4|j|)^{1/2}}(\zeta-\zeta_0)\right)\,,
\nonumber \\
 \zeta-\zeta_0 =n-n_0-v(\tau-\tau_0)\,,
\end{eqnarray}
corresponding to the tunneling between two adjacent levels.
There, a multitude of true dynamical level avoidings results in a 
system of
{\it kink trains} between subsequent pairs of levels. The limiting
velocity $ |v|=2\sqrt j$ corresponds to the limit of vanishing
fluctuation $\bar \delta_{2n+1} $ and thus to suppressing the
tunneling.

To consider now a situation interpolating between the two previous 
situations  we
take $\delta_{2n}= \bar\delta_{2n}+x$ in Eq.(\ref{11a}) for
$\delta_{2n}$, therefore choosing $x$ so as to eliminate the
quadratic terms in Eq.(\ref{2n}). Let us take approximately
$\delta_{2n+1}\approx\delta_{2n-1}\approx - \delta_{2n}$. Then we
arrive at equation
\begin{equation}
\frac{1}{4}\frac{\partial^2\bar\delta_{2n}}{\partial \tau^2}-n
\frac{\partial^2\bar\delta_{2n}}{\partial n^2} - \left (
\frac{2}{3}n-2|j| \right)\bar\delta_{2n}
 +2n \bar\delta_{2n}^3= \frac{4}{3}(2|j|-n/9) \,.
\label{Lal:eq}
\end{equation}
If the oscillations in $n$ can be neglected, under conditions for
the coefficients of Eq.(\ref{Lal:eq}), $B\equiv 2n/3-2|j|>0 $ and
$F\equiv (4/3)(2|j|-n/9) >0$ (or $n>3|j|> n/6$), there exists an
exact solution \cite{Lal:1986} to (\ref{Lal:eq}) of periodic
non-sinusoidal form spanned by $F$ which describes the kink
nucleation fluctuation. Increasing $j$ causes the growth of the energy
of the nucleus until a new kink generates. The respective solution reads
\begin{equation}
\bar\delta_{2n} (\tau)= a\frac{n_1+\cos(2 w \tau)}{n_2+\cos(2 w
\tau)},
\end{equation}
where $a, w, n_1$, and $n_2$ are given as
\begin{eqnarray}
n_1=(2-n_2^2)/n_2\,,\quad a^2= \frac{B}{2n} \frac{n_2^2}{2+n_2^2}\,,
\nonumber \\ 4 w^2 = 2 B \frac{n_2^2-1}{n_2^2+2}\,, \quad 
 a= -\frac{F}{2
B}(2+n_2^2).
\end{eqnarray}
From the relation $F= a(2na^2-B)$ it is evident that the amplitude
$a$ of the fluctuation $\bar\delta_{2n}$ is spanned by the driving
field $F$, growing with $|j|$ until the energy of the fluctuation
reaches the energy of the kink.

Equations  (\ref{2n}) and  (\ref{11a}) also admit a chaotic regime
for sufficiently large $j, n $ under conditions
 $\partial^2\delta_{2n}/\partial \tau^2=8|j-1/2|+c_1$
and
 $\partial^2\delta_{2n+1}/\partial \tau^2=-8|j-1/2|+c_2$,
where $c_1= c_2=0$, if the oscillations in $n$ can be neglected--
i.e. $\delta_{2n+1}+\delta_{2n-1}-2\delta_{2n}\approx \partial^2
\delta_{2n}/\partial n^2 = 0$ and the same for
 $\delta_{2n+1}$.
 One obtains then $\delta_{2n}+\delta_{2n+2}=
2n^2\bar\delta_{2n+1} (1-\bar\delta_{2n+1})/[|j-1/2|(n+j-1/2)]$, where
$\bar\delta_{2n+1}\equiv |j-1/2|\delta_{2n+1}/n.$ Since
$\delta_{2n+2}\approx \delta_{2n}$, the famous logistic equation is
recovered
\begin{equation}
\delta_{2n+2}\approx A \bar\delta_{2n+1}(1-\bar\delta _{2n+1}),
\label{log}
\end{equation}
where $A= n^2 /[|j-1/2|(n+j-1/2)]$.  Equation (\ref{log}) assumes the
transition to the chaotic regime for $A\geq A_{crit}=3.56994 \dots$,
i.e., for
\begin{equation}
n_{crit} = |j-1/2|\frac{A_{crit}}{2}\left
(1+\sqrt{1+\frac{4}{A_{crit}}}\right )=2.1921\cdot |2j-1|.
\label{crit}
\end{equation}
The condition $n>n_{crit}$  specifies the values of $n=n_r$ from the
nonlinear region above the kink nucleation threshold $n>3|j|$ as
estimated above. On the other side the maximum value of $A$, $A=4$,
implies, $n_4= (2j-1)(1+\sqrt 2)$, which determines the upper limit
for $n$, $n<n_4$.

 Traditionally the term ``quantum chaos'' is used to denote
the traces of {\it classical} chaotic behavior at a quantum level.
As already mentioned in the Introduction, the classical counterpart
of the system under consideration cannot be defined uniquely. The
semiclassical approximation in two-level systems generically leads
to classical chaotic patterns as a result of the nonlinear coupling
between two subsystems, boson and electron, considered, respectively,
as classical and quantum. The onset of classical chaos corresponds
to energies above the first diabatic line. Referring to Eq.
(\ref{crit}) it is to be emphasized that the chaotic
behavior refers to mentioned a purely quantum regime of 
medium values of $j$ and $n_r$
between the weak coupling with $j\ll n_r$ (dimerized pairs of
oscillators) and strong coupling with $j\gg n_r$ (kink lattice)
domains. Thus, this chaotic behavior can be regarded as being of
essentially quantum nature. One can conclude that the mapping of
a quantum system onto the classical Calogero-Moser gas with
repulsive interactions enables one to use the classical formalism
for describing the system via its quantum numbers. So a promising
feature of this approach is its ability to represent quantum chaos
by means of classical equations.

\section{Level statistics}

Statistical methods of investigation of the complex spectra are
based on evaluation of various "chaoticity degrees" of quantum
systems -- i.e., the density of levels, distribution of 
nearest-neighbor spacings (NNS's), and statistics of distribution of
"curvatures" accounting for avoided crossings as well as more
complicated measures of the distribution of avoided crossings
\cite{Brody:1981,Zakrzewski:91,Gaspard:1990}. Most of these
characteristics can be obtained within the framework of
random-matrix theory (RMT)
\cite{Brody:1981,Dyson:1962,Dyson:1972,Mehta:1960}, whose
predictions are supposed to adequately describe the quantum
systems with generically chaotic classical analogs. However, the
predictions of RMT cover in a satisfactory fashion only the
"chaotic" limit of these systems, classifying them into three
universality classes according to the time-reversibility of the
underlying Hamiltonian \cite{Dyson:1962,Mehta:1960,Brody:1981}
(that is, Gaussian orthogonal GOE, unitary GUE, and symplectic GSE
ensembles). The fourth universality class is sometimes marked as
corresponding to regular dynamics-- for example, if the classical
analog of the system for some values of controlling parameters
exhibits mostly regular KAM trajectories.

\begin{figure*}[th]
\includegraphics[scale=0.45]{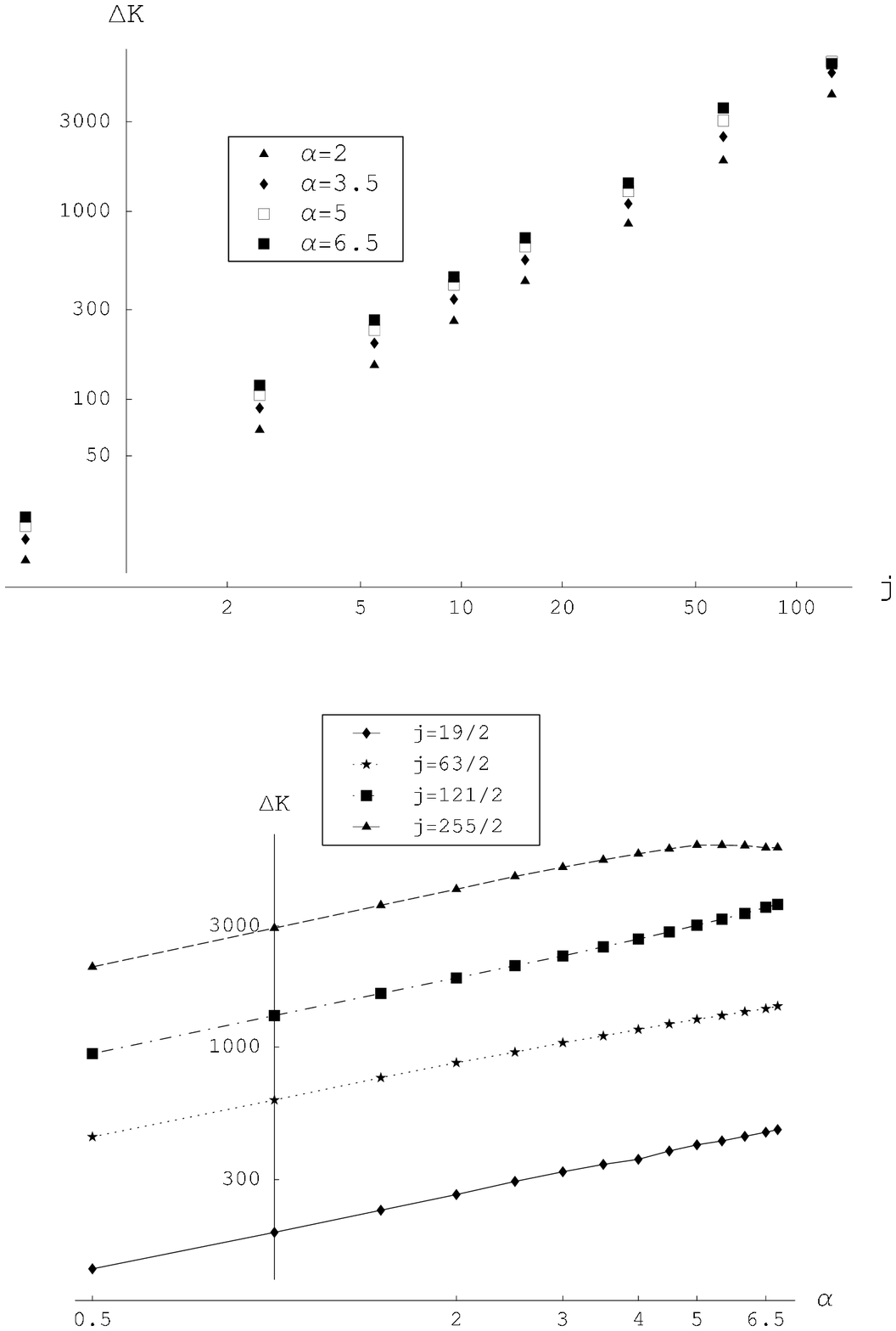}
\caption{The widths of curvature distributions $\Delta K\equiv
\langle \Delta^2 K \rangle $ as function of $j$ (a) and $\alpha$
(b). The scaling laws $\Delta K\sim j^{\nu}$ and $\Delta K\sim
\alpha^{\mu}$ are seen.}
 \label{Fig5-S2}
\end{figure*}

In recent years, however, there came an understanding that chaotic
patterns are to be sought not only from the energy-level
distribution themselves but rather from pictures of their
changing with the change of controlling parameters in the Hamiltonian
($\alpha$). Thus, an alternative statistical approach is based on
the concept of fully integrable description of the spectra by the
dynamical model of a Calogero-Moser gas of interacting (repulsing)
pseudoparticles \cite{Pechukas:1983,Gaspard:1990} where the
controlling parameter is considered a pseudo-time (Section IIIC).
The dynamical equations for the eigenfunctions $|x(\tau)\rangle$ of
this Hamiltonian bring, for example,  to the notion of level
curvature $\ddot{x}(\tau)$ \cite{Gaspard:1990,Zakrzewski:91} and
several conjectures about its relation to the nature of the underlying
RMT ensemble. Various characteristics of the level curvature
distributions were reported as handy indicators of quantum chaos
traces in a system. In particular, it was suggested
\cite{Nakamura:1985} that for energy regions with quantum chaos
there holds the scaling $\Delta K \sim j^\nu $ with positive $\nu$;
meanwhile, for nonchaotic regions $\nu$ is rather negative or close
to zero. We investigated the widths $\Delta K$ of the distributions
of the level curvatures P(K) displayed as functions of $j$ and 
$\alpha$ [Fig. 5(a), 5(b)]. The level statistics of Fig. 5 (and of the
following Fig. 6 and 7) was taken from the energy intervals above the
first diabatic line where multiple avoided crossings come into
being. For the improvement of statistics we used the standard recipe of
collecting the level sequences from a number of $\alpha$ values
lying in the neighborhood. Fig. 5(a) apparently shows the scaling
behavior $\Delta K \sim j^\nu$ with the scaling coefficient $\nu \in
(0.99-1.01)$ for all $\alpha$. In Fig. 5(b) we observe an another
interesting regularity: namely, the scaling law $\Delta K \sim
\alpha^{\mu}$ with $\mu\simeq 0.5\pm 0.05$. This latter observation
opens a challenging suggestion about the diffusive character of
spreading the probability function $P(K)$ in pseudotime $\alpha$.
From a closer inspection of these distribution functions one can even
anticipate that the resulting diffusive equation would have a
telegraph term and possibly is capable of bearing anomalous
diffusion patterns. We are going to return to this point elsewhere.

\begin{figure*}[th]
\includegraphics[scale=0.45]{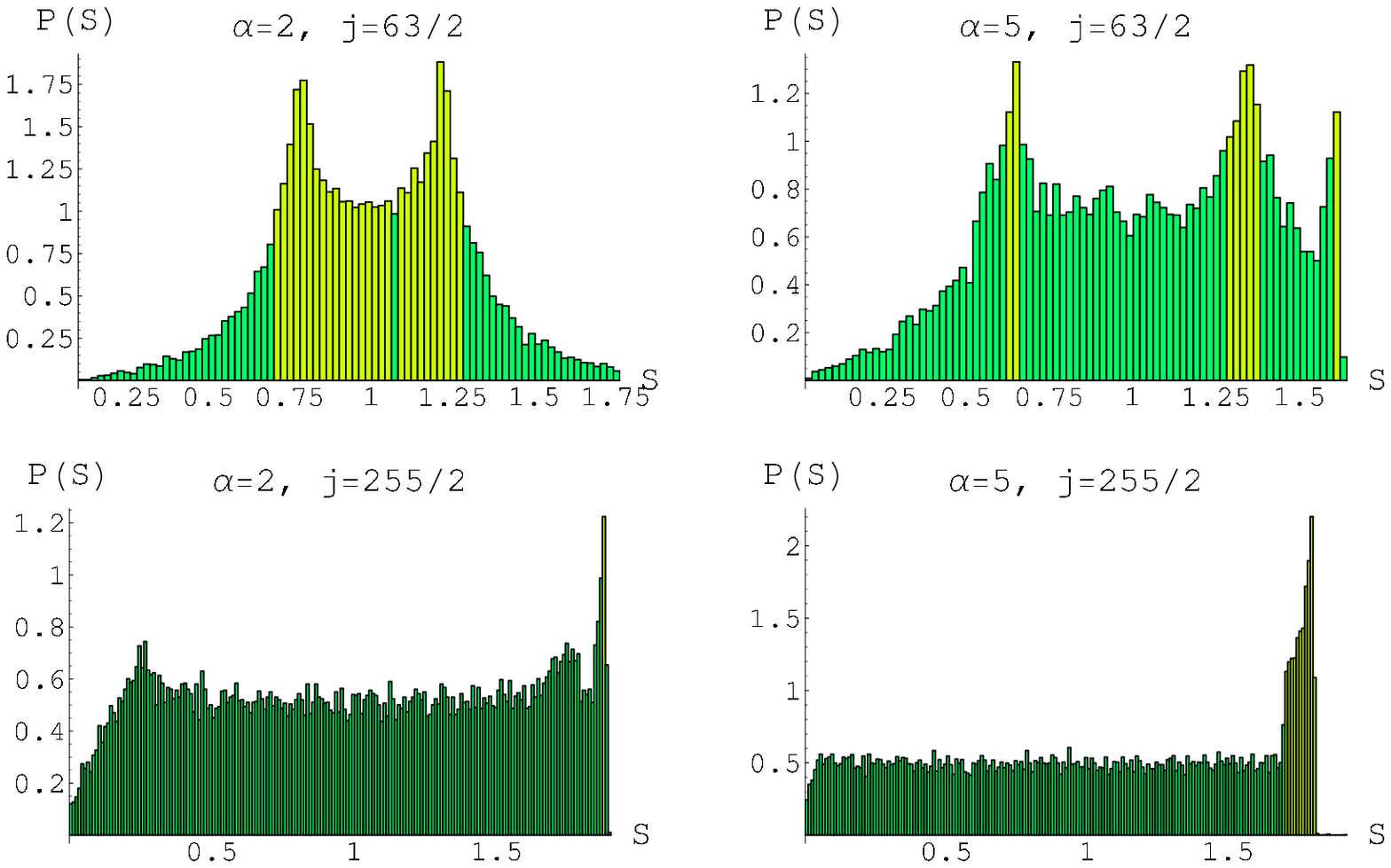}
\caption{(Color online) Level spacing distributions for particular
$j$. Dimerization clustering related to two dominating level
spacings is evident for $\alpha=2$ and $j=63/2$. The third peak from
localized states occurs at large $j$ and $\alpha$, dominating close
to the semiclassical limit (compare with Fig. 4). The crossover from
the dimerized regime to the localized regime with increasing $j$ and
$\alpha$ is evident.} \label{FigS2-b}
\end{figure*}

The realm of the symmetric JT model with definite $j$ value is that
of an effectively two-dimensional oscillator (or rather two oscillators
pertaining to two electron levels); thus, the level-spacing picture
for the unperturbed problem $\alpha=0$ is trivial: all levels of a
quantum harmonic oscillator are equidistant, so $P(S)=\delta(S-1)$.
The nontrivial behavior of the NNS distribution appears to be
consequence of the effective coupling of two oscillators through $\alpha$.
In Fig. 6 we show the level-spacing distributions $P(S)$ for a set
of $j$ with different $\alpha$. Our results indicate that the most
close to RMT prediction situations [RMT would yield a Wigner
distribution of NNS's in the form $P(S) \sim S \exp(-S^2)$]
 are located at intermediate values of $j$
and $\alpha$ as it is seen for pictures with $j=63/2$. Meanwhile, the
extreme values of $j$ and $\alpha$ do not fit RMT at all.
However, one can notice a crossover between two distinctly different
types of behavior seen, for example, at $j=63/2, \alpha=2$ and
$j=255/2$. Such a behavior can be semiquantitatively explained on
the basis of the conception of effective potential wells-- for
example, built on the semiclassical coherent probe function in the
space of their parameters (see \cite{Majernikova:2003}; this variant
of building an effective potential is in fact the Husimi
representation of the Hamiltonian operator taken for zero values of
classical momenta). Namely, the characteristic for big $j$ is the
presence of two almost noninteracting wells which are seen from the
right-hand of Fig. 4 showing spectral entropies. We can assume that each
of the wells refers to a separate quantum oscillator having level
spacing $a$ and $b$ (let $a < b$). The superposition of the distribution
for two independent sets of levels is found to be in the form (see
for details, e.g., Berry et al\cite{Berry:84}):

\begin{equation}
P(S)\simeq \frac{2}{b}\theta(a-S) + \delta(S-a) \,, \label{Ps:wells}
\end{equation}
$\theta$ being step function [we considered $a \ll b$; if they are
comparable, a refinement to (\ref{Ps:wells}) is straightforward].
Conformity of Fig. 6
 to the prediction (\ref{Ps:wells})
with $a \simeq 1.7$ and $b=4$ for large $j$ and $\alpha$ (for
$j=255/2$, $\alpha\geq 5$) is perceivable. From the ``adiabatic
sheet'' treatment in \cite{Majernikova:2003} one can even define
the parameters of both effective potential wells. The opposite
case of small $j$ ($\alpha$) apparently indicates a dimerized
structure with two dominating spacings, with increasing extent of
correlation when decreasing $j$ and $\alpha$.
 The graphs of the M-shaped probability distributions for fixed $j$
 in Fig. 6(a) for small $j$ and large $n$ are similar to those by Cibils
 {\it et al.} \cite{Cibils:1995} for $s=1/2$ which corresponds 
to our parity $p=1$.
 There, the asymmetry
between the matrix elements $f_{nn}$ and $f_{n+1,n}$ is negligible
and the weak-coupling approach is satisfactory. However, for large
$j$ with large asymmetry of the matrix elements the strong-coupling
limit of kink lattice is valid. There the relief of the three-minimum
ground-state potential unfolds (Figs. 4) and  the three-peak picture
arises [Figs. 6(b) and 6(c)].

\begin{figure*}[bh]
\includegraphics[scale=0.55]{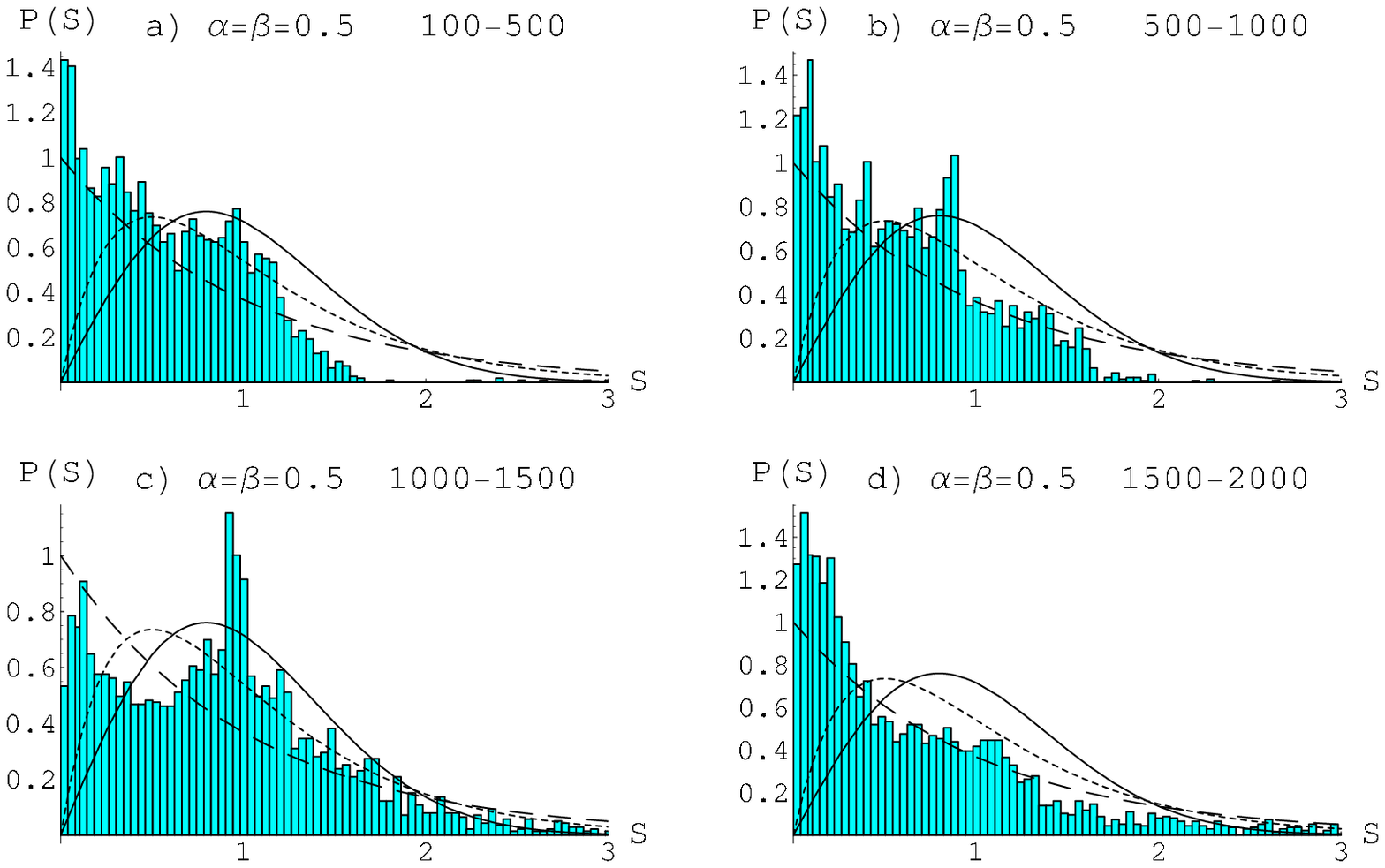}
\caption{(Color online) Level-spacing distributions for subsequent
segments of cumulative energy spectrum for small $\alpha<1$.
Dimerization clustering peaks as remnants of the distribution in
Fig. 6 are evident. The reference curves pertain to Wigner (solid line), semi-Poisson (dotted line),
 and Poisson (dashed line) distributions.
\label{Psa05b05}}
\end{figure*}

The level-spacing statistics taken separately for particular values
of $j$ is the basis of a more realistic investigation of the level
statistics of the whole spectrum of the system whose Hamiltonian in
this representation is a block matrix of blocks on the diagonal
corresponding to definite $j$ and off-diagonal blocks coupling [in
the generalized E$\otimes(b_1+b_2)$ JT model] for different $j$'s. The
peculiarity of the symmetric Jahn-Teller system is that the
interblock coupling is exactly zero, and as was shown already in
\cite{Rosenzweig:60} with a variant of the central limit theorem,
the superposition of (infinite or very large number) independent
blocks will give the cumulative distribution which must be of
Poissonian character, $P(S)= \exp (-S)$. This result is exactly that
obtained by Yamasaki {\it et al.} \cite{Nakamura:2003} for the absence of
(trigonal) nonlinearity and which was claimed to be testimony of the
absence of any traces of quantum chaos in this case. The detailed
account for the level statistics for the generalized JT model
including the symmetric JT as its partial case is reported by us
elsewhere \cite{Majernikova:2005}. For the sake of the present paper
we just mention the serious deviations of the expected Poisson
result in the symmetric case in the domain of dominating quantum
fluctuations ($1 > (\alpha, \beta)$). Indeed, the $P(S)$
distribution shown for $\alpha=\beta=0.5$ exhibits considerable
deviations from the expected Poisson shape, showing anomalously large
variance $>4$ (for $\alpha=\beta=1$ this dispersion is $\simeq
1.17$, and for the domain $1 <\alpha$ it is rather rigid, falling
normally within $0.7-0.8$ irrespectively to the values of $\alpha$).
These anomalies are apparently due to the high nonuniformity of the
spectral statistics in this case for different energy intervals. In
Fig. 7 we show the level-spacing distributions for
$\alpha=\beta=0.5$ taken for separate level intervals ($100-500$,
$500-1000$, $1000-1500$, and $1500-2000$) from which this difference
is apparent. The distributions with $\alpha,\beta >1$, in contrast,
are rather robust with respect to changing energy intervals.

\section{Conclusion}

The energy spectrum of a long but finite chain of correlated clusters is
governed by the interplay  (competition) of intracluster and intercluster
interactions (repulsions) determined by two quantum numbers:
rotational $j$ and radial $n_r$.
 At small $|j|$,  $|j|\ll n_r$,
  the excited spectrum is  long-range-ordered and the broad
{\it  dimerized} region is a sample of pairwise oscillating levels.
Respective level avoidings in the upper part of the spectra
[the inset in Fig. 1(a)] result from the oscillations due to comparable
intracluster and intercluster interactions in this part of the spectra.
 Such a regular wavelike spectrum is known to be
 characteristic for one-boson two-level systems since long time
ago \cite{Kus:1985}
 and is a signature of the dimerization. The similarity of this part of
 the E$\otimes$e spectra with two bosons to the spectra of
one-boson systems was
 discussed in a more detail in the Introduction.

 In the strong coupling limit of
  large $j$, $|j|\gg n_r$ (the intercluster
interaction is small or negligible when compared to the intracluster
one) there appears a {\it kink lattice} -- i.e., the regular lattice of
oscillating level clusters. There the levels are bridged by a flip
(kink) up to a higher level due to the tunneling [Fig. 1(b) above the
diabatic line and inset therein]. At moderate values of $|j|$ and $\alpha$
there appears an intermediate region of strong  nonlinear fluctuations
responsible for the {\it nucleation of the kinks} [Figs. 1, 4(c),and 4(d)].

 By an approximate quantum
treatment of both the strong- and weak-coupling
 limit in the Sec.III we have found analytically the respective
wavefunctions and energies of
localized exotic and extended states in a form of modified coherent
oscillators: at strong coupling we showed the appearance of two branches
of coherent oscillators with effective frequencies $\Omega^{(\pm
)}_{eff}\equiv  2 \pm \alpha/\sqrt{2j-1}$ squeezed or antisqueezed
by $j$ to the degenerate quantum limit $\Omega=2$ at $j\rightarrow
\infty$. An excellent agreement of these solutions to the
exact
numerical wave functions for large $j$ ($\geq 63/2$) was marked. In the 
weak-coupling limit of small $j \ll n_r$
the first order of perturbation theory yields satisfactory agreement
with exact results for the wave functions except for $\alpha \sim 1
$ ($\Omega\sim 1$). Therefore simple
 approximate analytical solutions for wave functions are available in
both the strong- and weak-coupling limits corresponding to the soliton
(kink) and wavelike lattice (in the plane energy-coupling
parameter).

 The mapping of our model on the Calogero-Moser gas of
classical pseudoparticles with repulsive interactions enabled us to
describe (Sec.III C) all regions of the spectra by a set of
classical equations in terms of its quantum numbers. In the intermediate
region the equations under certain conditions implied the
logistic equation route to chaos
in terms of quantum numbers.

 On the other hand, the qualitative
features of the excited spectra of the E$\otimes$e  Jahn-Teller model
can be understood  from the outline of the shape of the effective
potential  used for investigating the ground state of the
model \cite{Majernikova:2003,Majernikova:2002}. With increasing $j$
and $\alpha$ up to the semiclassical limit there emerge up to three
effective potential minima. In accordance with this we identify the
above-mentioned three phases in the spectra: for small coupling
$\alpha $ the dimerized phase is related to two broad minima of the
potential. Increasing $\alpha $ opens one additional narrow minimum
(for the ground state its counterpart was responsible for emerging
the "tunneling phase" or light polaron \cite{Majernikova:2002}). The
complex interplay of wave functions located over three minima
corresponds to our intermediate region [Fig. 4(c) and 4(d)] where the level
statistics (Fig. 6) exhibits chaotic patterns close below the limit
expected for fully chaotic RMT statistics \cite{Dyson:1962}. Next,
subsequently increasing $\alpha$ and $j$ brings a suppression of the
initial two "weak-coupling" wide minima in the quantum limit of
large $j$. Thus there persist two almost noninteracting potential
wells: a narrow and a wide one. Two remaining noninteracting
branches of the spectra are markedly seen in Fig. 4(f). The lowest
branch of the spectral entropy in Fig. 4(f) is related to the (third)
narrow minimum of the effective potential responsible for the kink
lattice phase of avoided crossings (localized states) in energy
spectra. This behavior is also supported by the shape of the 
level-spacing distribution [Fig. 6(d) and Eq.(\ref{Ps:wells})] which has
the marked character of superimposing two almost noninteracting
effective oscillators with different frequencies.

The statistical level-spacing probability distributions investigated
in Sec.IV were numerically calculated for particular $j$'s.
Their features are in agreement with above conclusions based on the
shape of the effective potential:  They are shown to exhibit up to three
maxima (Fig. 6) depending on $j$ and $\alpha$. For moderate $j$ and
$\alpha$ a
 pair of almost symmetric prominent peaks of $P(s)$ are related to two
dominating strongly correlated potential wells which are getting
closer at small $\alpha $ and $j$. The two dominating peaks of level
spacings refer to the dimerization clustering of subsequent even and
odd levels.
 At large $j$ and $\alpha$
 the NNS distributions are strictly determined by the shape of the
ground-state effective potential (Fig. 6(d). The analysis of the cumulative
statistical distributions of level spacings shows a strongly changing
picture of different segments of spectra especially at weak
couplings (Fig. 7). The dimerized phase is pronounced in the lowest
part of spectra where a strong clustering is apparent from two peaks
of the level-spacing distribution $P(S)$. In the middle part of
spectra some patterns of the Wigner distribution at least at large $S$
are perceivable. The widths of the level curvature distributions were
found (Fig.5) to be scaled as $\Delta K\propto j^{\nu}$, $\nu\propto
1$, and thus satisfying the criterion suggested \cite{Nakamura:1985}
for indicating chaotic patterns in our quantum system. Additionally,
the scaling behavior $\Delta K\propto \alpha^{0.5}$ was revealed
which traces a possible bridge between mechanical and statistical
points of understanding the complexity in the system referring to
a possibility of reformulating the level curvature statistics
stochastically, as a Calogero-Moser gas diffusing with "time"
$\alpha$ \cite{Hasegawa:1988}.

Since the sets of levels with $j\neq j^{'}$ for the symmetric JT
system are uncorrelated (matrix elements between the states are
zero), the respective levels intersect (Sec. II). The cumulative
level-spacing probability distributions consist of independent
contributions of all $j^{'}$s and, consequently, the Poissonian
distribution should supercede
 the above-described  partial distributions. The cumulative
level-spacing distributions, however, are found to exhibit
enormously large dispersions $\gg 1$ at small $\alpha \leq  1$
(Fig. 7) which can be  evidence for the overlapping of
distributions of two kinds: a Poissonian one from the crossing of
uncorrelated levels $j\neq j^{'}$ and two-peak ones for the
particular $j$'s described above. In the quantum limit
$j\rightarrow \infty$ the spectrum tends to a continuum and the
respective levels  at $j\rightarrow j^{'}$ will tend to avoiding
levels.

\vspace{0.5cm}

{\bf Acknowledgement}. It is a pleasure to thank Professor J. Pe\v
rina for useful discussion. This research was partially supported
by Project No. 202/06/0396 of the Grant Agency of the Czech
Republic as well as by Project No. 2/6073/26 of the Grant Agency
VEGA of the Slovak Academy of Sciences.  One of us (S.Sh.) is also
indebted for partial support by Project No. LN00A015 of the
Ministry of Education of the Czech Republic as well as to the
Department of Theoretical Physics of the Faculty of Science of the
Palack\'y University in Olomouc.


\begin{thebibliography}{99}
\bibitem{ftn}Note that the
numeration of states in Fig. 3 labels the states with given $j$, and
in Fig. 2 all states are labeled in ascending order in energy (that is
contain all $j$).
\end{thebibliography}
\end{document}